# On-chip random spectrometer

Brandon Redding, Seng Fatt Liew, Raktim Sarma, Hui Cao*

Department of Applied Physics, Yale University, New Haven, CT 06520, USA

*E-mail: hui.cao@yale.edu

Light scattering in disordered media has been studied extensively due to its prevalence in natural and artificial systems [1]. In the field of photonics most of the research has focused on understanding and mitigating the effects of scattering, which are often detrimental. For certain applications, however, intentionally introducing disorder can actually improve the device performance, e.g., in photovoltaics optical scattering improves the efficiency of light harvesting [2–5]. Here, we utilize multiple scattering in a random photonic structure to build a compact on-chip spectrometer. The probe signal diffuses through a scattering medium generating wavelength-dependent speckle patterns which can be used to recover the input spectrum after calibration. Multiple scattering increases the optical pathlength by folding the paths in a confined geometry, enhancing the spectral decorrelation of speckle patterns and thus increasing the spectral resolution. By designing and fabricating the spectrometer on a silicon wafer, we are able to efficiently channel the scattered light to the detectors, minimizing the reflection loss. We demonstrate spectral resolution of 0.75 nm at a wavelength of 1500 nm in a 25 μm by 50 μm random structure. Furthermore, the phenomenal control afforded by semiconductor nanofabrication technology enabled us to engineer the disorder to reduce the out-of-plane scattering loss. Such a compact, high-resolution spectrometer that is integrated on a silicon chip and robust against fabrication imperfections is well suited for lab-on-a-chip spectroscopy applications.

Spectrometers are widely used tools in chemical and biological sensing, material analysis, and light source characterization. The development of a high-resolution on-chip spectrometer could enable compact, low-cost spectroscopy for portable sensing as well as increasing lab-on-a-chip functionality. However, the spectral resolution of grating-based spectrometers scales with the optical pathlength, which translates to the linear dimension or footprint of the system. As a result, on-chip spectrometers based on curved gratings (Echelle) [6,7] and arrayed waveguide gratings [7–11] require relatively large footprint (~cm). This limitation inspired researchers to develop a number of alternative spectrometer designs. On-chip digital planar holography [12–14] has been shown to provide high resolution, but the sensitivity is limited. A dispersive photonic crystal lattice [15], operating in the slow light regime, combines high resolution with small footprint; however, it has only been applied to the detection of individual spectral lines. Resonant devices such as microrings [16–18], microdonuts [19], and photonic crystal defect cavities [20] make the effective interaction length much longer than the physical dimension of the device, thus providing high resolution in a small footprint; unfortunately, these devices are particularly sensitive to fabrication errors.

In addition to regular systems, disorder and scattering have also been explored for spectroscopy applications. Xu et al. used spatio-spectral transmission patterns of disordered photonic crystals to construct multimodal spectrometers [21]. Kohlgraf-Owens and Dogariu showed that random scattering materials have sufficient diversity in spectral transmission to allow for precise measurements of the spectrally dependent polarization state of an optical field [22]. The working principle of random spectrometers is that the speckle pattern formed by transmitted light through a disordered system provides a sort of fingerprint, uniquely identifying the wavelength of the probe signal. In practice, the wavelength-dependent speckle patterns are measured and stored in a transmission matrix, which describes the spectral-to-spatial mapping of the spectrometer. After calibrating the transmission matrix, an arbitrary input spectrum can be reconstructed from its speckle pattern. This approach has also been applied to build spectrometers with an array of Bragg fibers [23], or a single multimode fiber [24,25]. The advantage of utilizing multiple scattering in a disordered medium is that it folds the optical paths, making the effective pathlength longer than the linear dimension of the system. Thus a small shift in the input wavelength will cause a significant change in the transmitted speckle pattern. In other words, multiple scattering enhances the spectral decorrelation of speckle patterns, enabling fine spectral resolution with a limited footprint. For instance, in the diffusive regime, the effective optical pathlength scales as the square of the actual length, $L$, of the system [26]. Thus the spectral resolution, which is determined by the spectral correlation width of transmitted speckle, scales as $1/L^2$ instead of $1/L$. This enhancement occurs at all frequencies, unlike the resonant cavities which enhances the optical pathlength only at discrete frequencies. However, the total transmission through a diffusive system of length, $L$, and transport mean free path, $l_t$, is approximately $l_t/L$. When $L$ is much larger than $l_t$, most of the input signal is reflected instead of being transmitted. This loss will limit the spectrometer sensitivity.

In this work, we present the first demonstration of an on-chip spectrometer based on multiple scattering in a disordered photonic structure. The increased optical pathlength enabled fine spectral resolution in a small footprint. Furthermore, the control afforded by designing an on-chip spectrometer allowed us to mitigate the high insertion loss normally associated with random scattering media. By surrounding the random structure with a full-bandgap photonic crystal boundary, we efficiently channeled the diffusive light through the disordered medium to the detectors. We also tailored the scattering properties of the random system, which consisted of precisely positioned air cylinders etched into a silicon membrane. By introducing structural correlations to the disordered medium, we engineered the spatial Fourier spectra to reduce the out-of-plane scattering loss.

We designed and fabricated the random spectrometer in a silicon-on-insulator (SOI) wafer. As shown in the scanning electron microscope (SEM) images in Fig. 1(a), the two-dimensional (2D) scattering structure is a random array of air holes etched into the silicon layer. A ridge waveguide delivered the probe light to the random array, where light was scattered by the air holes and began diffusing in all directions. The signal that reached the other end of the random structure was detected. In order to maximize the number of detectors and ensure that the physical distance from the input end to each detector was constant, we patterned the air holes in a semicircle. The probe signal entered from the center of the semicircle, and diffused outward until reaching the edge of the circle. The intensity distribution along the edge of the semicircle was used as the "fingerprint" to uniquely identify the input spectrum. To eliminate the loss due to light escaping from the base (straight segment) of the semicircle, we placed a photonic crystal layer (periodic array of air holes) with a full bandgap along the base. A row of holes was removed to create a defect waveguide for the input light. Similar photonic crystal boundary and defect waveguides were introduced along the circumference of the semicircle. The multiply scattered light that reached these waveguides was channeled to the detectors. The light that hit the photonic crystal layer in between the waveguides was reflected back to the random structure and went through further scattering until arriving at one of the defect waveguides. The output waveguides were separated by five rows of the triangular lattice of air holes to minimize their coupling. The width of each waveguide was tapered to match the size of the detector at the end. To avoid the complexity of integrating detectors in the proof-of-concept demonstration, we terminated the output waveguides by a semicircular ridge which scattered light out of the plane. The intensity of scattered light is proportional to that collected by each waveguide, and we imaged the scattered light from the top with a camera. A representative image of the scattered optical signal is shown in Fig. 1(c). The input light was provided by a laser operating at $\lambda = 1500$ nm. The intensity of light coupled to each output waveguide was extracted by integrating the scattered intensity in each detector region, as outlined by the white lines in Fig. 1(c). We patterned the random spectrometer by electron beam lithography and etched in an inductively coupled plasma reactive ion etcher (see the Methods). The scattering strength was controlled via the size and density of air holes. To model the random spectrometer, we performed numerical simulations using the finite difference frequency domain (FDFD) method. Figure 1(b) shows the $H_z$ field amplitude of TE polarized light diffusing through the semicircular random structure and coupling to the output waveguides.

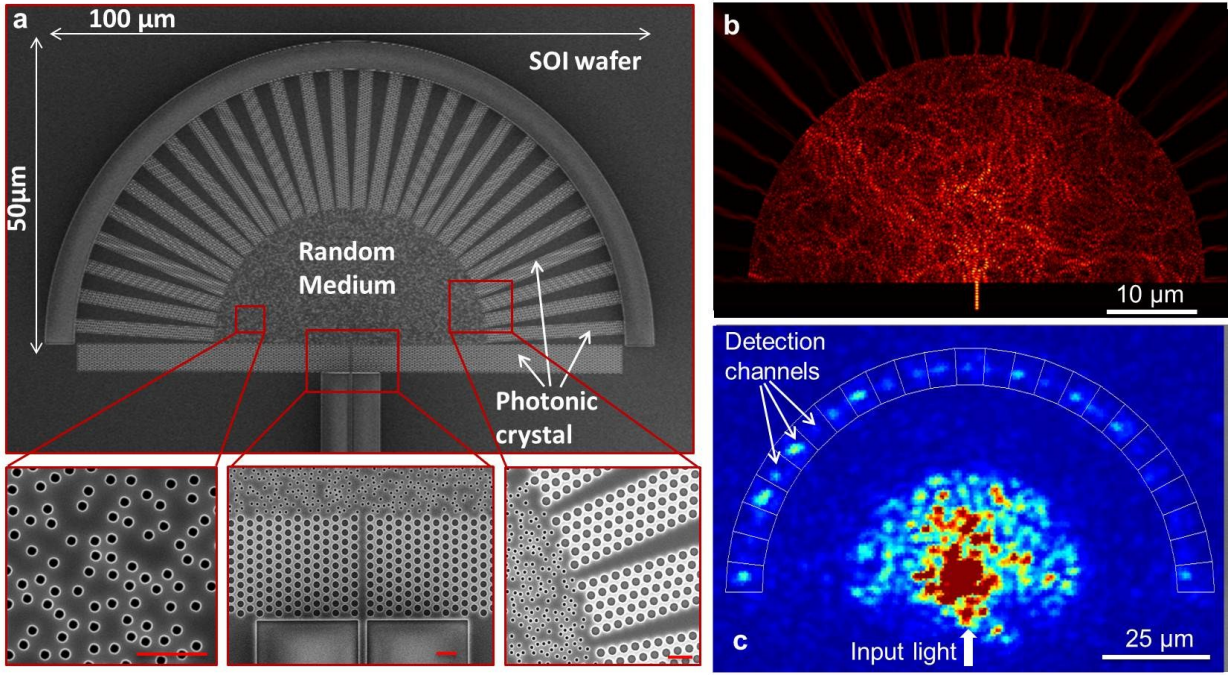

**Figure 1 | An on-chip spectrometer based on multiple scattering in a disordered photonic structure.** (a) Scanning electron microscope image of the fabricated spectrometer. The dispersive element is a semicircular array of randomly positioned air holes, surrounded by a photonic crystal lattice. The probe signal is coupled to the random structure via a defect waveguide at the bottom of the semicircle. The light then diffuses through the random array via multiple scattering and eventually reaches the 25 defect waveguides around the circumference of the semicircle. These tapered waveguides will couple the signals to the detectors (not integrated). The distribution of intensities over the detectors is used to identify the input spectrum. The photonic crystal boundary, which has a full bandgap in 2D, confines the probe light in the random structure and channels it efficiently into the defect waveguides. The insets in the bottom row are magnified images, and the scale bars indicate 1 μm. (b) Numerical simulation of TE polarized light at $\lambda$ = 1500 nm diffusing through the random structure. The amplitude of the Hz field shown here is calculated by the finite-difference frequency-domain method. (c) Experimental near-infrared optical image of the random spectrometer with a probe signal at $\lambda$ = 1500 nm. The white boxes, labeled "Detection channels", mark the positions of detectors at the end of 25 defect waveguides. To avoid the complexity of integrating the detectors, we estimated the intensity coupled into each output waveguide from the integrated intensity of scattered light within each white box. The out-of-plane scattering is caused by the semicircular groove, shown in (a), that terminates the waveguides at the location of the detectors.

The spectral resolution of the random spectrometer depends on the change in wavelength required to generate an uncorrelated intensity distribution on the detectors. It can be quantified by the spectral correlation function of the intensity on the detector plane as: $C(\Delta\lambda, x) = \langle I(\lambda, x) I(\lambda+\Delta\lambda, x) \rangle / [\langle I(\lambda, x) \rangle \langle I(\lambda+\Delta\lambda, x) \rangle] - 1$, where $I(\lambda, x)$ is the intensity recorded by detector $x$ for input wavelength $\lambda$, and $\langle \cdots \rangle$ represents the average over $\lambda$. We measured $I(\lambda, x)$ by recording images such as the one shown in Fig. 1(c) as a function of probe wavelength. Then the spectral correlation function was computed and averaged over all detectors, as shown in Fig. 2(a) for a random spectrometer of 25 μm radius. $C$ is normalized to 1 at $\Delta\lambda = 0$, and its half width at half maximum, $\delta\lambda$, is 0.6 nm, meaning that a wavelength shift of 0.6 nm is sufficient to reduce the degree of correlation of the speckle pattern to 0.5. $\delta\lambda$ provides an estimate of the spectral resolution, because it is impossible to resolve two wavelengths with highly correlated speckle patterns. The actual resolution also depends on the reconstruction algorithm and the experimental noise of the measurements.

To use the random system as a spectrometer, we first calibrated the spectral-to-spatial mapping by recording the wavelength-dependent intensity distributions on the detectors. This calibration was stored in a transmission matrix, $T$, relating the discretized spectral channels of input, $S$, to the intensity measured by different detectors, $I$, as $I = T S$ [25]. Each column in $T$ describes the intensity distribution on the detectors produced by input light in one spectral channel. The number of independent spectral channels (separated by $2\delta\lambda$) that can be measured simultaneously is limited by the number of independent spatial channels [25]. For a 25 μm radius spectrometer with 25 independent detectors, we chose a bandwidth of 25 nm: from $\lambda = 1500$ nm to 1525 nm. The spectral channel spacing was selected to be 0.25 nm, which is less than $\delta\lambda$ to test the limit of the spectrometer resolution. Calibration was then conducted by setting a tunable laser to the center wavelength of each spectral channel and recording the intensity distribution, thereby measuring $T$ one column at a time. A representative transmission matrix is shown in Fig. 2(b).

After calibration, an arbitrary probe spectrum can be reconstructed by measuring the intensity of light reaching the detectors ($I$) and multiplying it by the inverse of the transmission matrix: $S = T^{-1}I$. In practice, the matrix inversion process is ill-conditioned in the presence of experimental noise. To mitigate the effects of noise, we used a truncated inversion algorithm based on singular value decomposition, followed by a nonlinear optimization procedure to find the input spectra $S$ that minimizes $\|I-TS\|^2$ [25]. Using this combination of truncated inversion and nonlinear optimization, we tested the ability of the random spectrometer to resolve a series of narrow lines across the 25 nm bandwidth. As shown in Fig. 2(c), the spectrometer accurately recovers the positions of the individual lines with an average signal-to-noise ratio of ~1000. The linewidth is less than 0.5 nm. We then characterized the spectral resolution of the spectrometer by testing its ability to resolve two closely spaced spectral lines. To synthesize the probe spectrum, we added the intensity recorded separately on the detectors by the two spectral lines: $I_{probe} = I_{\lambda 1} + I_{\lambda 2}$, since light at different wavelengths does not interfere. As shown in Fig. 2(d), two lines separated by merely 0.75 nm are clearly resolved. This confirms that multiple scattering in a disordered medium enables sub-nm spectral resolution with a 25 μm by 50 μm footprint.

The above calibration and testing were done with TE polarized light (electric field parallel to the silicon layer). The same random structure can also function as a spectrometer for TM polarized light (electric field perpendicular to the silicon layer), as long as the transmission matrix for the TM polarization, which differs from that of TE, is calibrated. While the grating-based on-chip spectrometer works only for a fixed spectral range because the monolithic grating cannot be rotated, the random spectrometer can operate at varying spectral regions without structural modification. This is because multiple scattering occurs in a random structure over an extremely broad range of frequency. A switch of the operation frequency can be done simply by changing the transmission matrix to the one calibrated for the desired spectral region. However, care must be taken to ensure that no input signal outside the operation bandwidth is coupled to the spectrometer, as this would corrupt the spectral reconstruction.

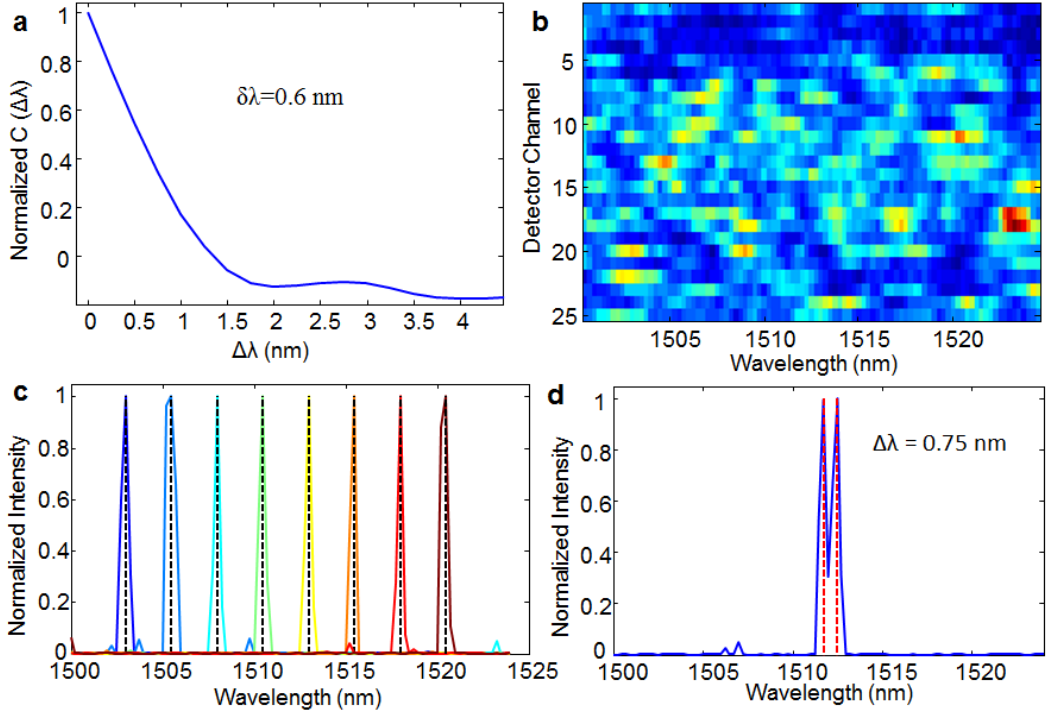

**Figure 2 | Spectral calibration and testing of the random spectrometer.** (a) The spectral correlation function of light intensities averaged over all detection channels of a 25 μm radius spectrometer. The half width at half maximum is 0.6 nm, meaning a wavelength shift of 0.6 nm reduces the degree of spectral correlation to half. (b) The transmission matrix stores the measured intensity distribution on the detection channels as a function of the input wavelength. The matrix was calibrated by recording images such as the one in Fig. 1(c) for each spectral channel with a wavelength tunable laser source. (c) Reconstructed spectra for a series of narrow spectral lines across the 25 nm bandwidth. The black dotted lines mark the center wavelength of each probe line. The width of each reconstructed line is less than 0.5 nm, and the average signal-to-noise ratio is over 1000. (d) Reconstructed spectrum (blue line) of two narrow spectral lines separated by 0.75 nm. The red dotted lines mark the center wavelengths of the probe lines.

In addition to spectral resolution, bandwidth, and footprint, sensitivity is another crucial metric of spectrometer performance. For the on-chip random spectrometer, a good sensitivity requires maximizing the transmission from the input waveguide through the scattering structure to the detectors. The low transmission normally associated with a disordered scattering medium is due to the open boundary: light can escape from the disordered medium in any direction. By surrounding the disordered structure with a reflecting photonic crystal boundary, we intended to confine light in the random system, limiting the escape routes to the defect waveguides which lead to the detectors. To estimate the transmission through the disordered medium in our spectrometer, we performed 2D FDFD simulations with and without the photonic crystal boundary (see Supplementary Information). The simulation results showed that the photonic crystal with a full bandgap dramatically improved the collection efficiency. For a semicircular random medium of 25 μm radius, 60% of the input light was channeled into the output waveguides with the remaining 40% returning to the input waveguide; without the photonic crystal boundary, only 21% of the input reached the detectors.

The 2D simulation, however, neglected loss due to out-of-plane scattering. In the near-field image of the random spectrometer [Fig. 1(c)], we observed a strong signal from within the random structure itself, indicating that a significant fraction of the input light was scattered out-of-plane before reaching the detectors. Note that the out-of-plane scattering limits not only the spectrometer sensitivity, but also the spectral resolution. For a random spectrometer of 25 micron radius, the experimentally measured spectral correlation width $\delta\lambda$ is 0.6 nm, while the 2D simulation of the same structure, ignoring the out-of-plane scattering, gives $\delta\lambda \sim 0.3$ nm. This is because the out-of-plane leakage is larger for the longer optical paths, thereby preferentially attenuating the light going through longer paths and reducing the effective pathlength of light reaching the detectors. Since the spectral resolution depends on the optical pathlength, the out-of-plane leakage limits the resolution.

To reduce the out-of-plane scattering, we note that it occurs when the scattering from the disordered media reduces the magnitude of the in-plane propagation constant, $k_{\parallel}$, of the light such that it is no longer confined outside the light cone ($|k_{\parallel}| > \omega/c$, $\omega$ is the angular frequency and $c$ is the speed of light). As the light index-guided in the silicon layer undergoes scattering, the in-plane propagation constant changes as $k'_{\parallel} = k_{\parallel} + q$, where $k'_{\parallel}$ is the new in-plane propagation constant and $q$ is the spatial vector of the scattering medium. As long as $|k'_{\parallel}| > \omega/c$, the scattered light remains outside the light cone. However, if $|k'_{\parallel}| < \omega/c$, the scattered light can leak out of the silicon layer into the air, reducing the collection efficiency of the spectrometer. By engineering the disorder, we can control the spatial vectors present in the scattering medium and influence the available $q$'s for out-of-plane scattering.

We therefore sought to replace the completely random structure with partially random ones by introducing structural correlations. In particular, we considered two alternative scattering media: a photonic amorphous structure and a golden-angle spiral lattice. The former [Fig. 3(b)] has short-range order [27–29], as there is a characteristic spacing of adjacent scatterers - air holes [30]. The latter [Fig. 3(c)] is a deterministic aperiodic structure [31,32], which has been used in the arrangement of seeds in sunflower heads to ensure the most even distribution of seeds without clumping [33]. We conducted the spatial Fourier transform of these two patterns to compare with a random pattern. The amplitude of the spatial Fourier spectra, plotted in Fig. 3(d-f), represents the likelihood of finding a spatial vector $q$. The random structure has all possible spatial vectors, and its Fourier spectrum is continuous. The photonic amorphous structure and the golden-angle spiral lattice, in contrast, exhibit bright circles, indicating the existence of dominant spatial vectors. If these vectors have large enough amplitudes, most scattering events will keep the light outside the light cone. Hence, by adjusting the characteristic spacing of air holes, we can lower the probability of out-of-plane scattering.

We designed the amorphous and spiral structures for the on-chip spectrometer. To operate at the wavelength of ~1500 nm, the average spacing of air holes was chosen to be 343 nm, and the radius of air holes was 75 nm. For comparison, we also made a random structure with the same size and density of air holes. To estimate the out-of-plane scattering loss, we performed numerical simulations of all three scattering media. The full 3D simulations are computationally heavy; we therefore calculated the fields in 2D using an effective index of refraction for the silicon layer (see Supplementary Information). We considered the TE polarized light with in-plane electric fields, and performed a 2D Fourier transform of the fields to obtain the wavevectors of light propagating inside the system. We then computed the fraction of wavevectors within the light cone in order to estimate the relative strength of out-of-plane scattering for the random, amorphous, and spiral structures [32]. Experimentally there are two light cones, one for air above the silicon layer, the other for silica underneath ($|k_{\parallel}| > n_s \omega/c$, $n_s = 1.5$ is the refractive index of silica). Since the latter is larger than the former, we used it in the computation, and found the light in the random medium had 38% and 81% more energy inside the light cone than the amorphous and spiral structures, respectively (see Supplementary Information). These results confirmed our expectation that the structural correlations can be used to reduce the out-of-plane scattering.

Finally, we fabricated a set of spectrometers with all three scattering media to perform an experimental comparison. SEM images of the three scattering media are shown in Fig. 3(g-i). We monitored the out-of-plane scattering by imaging the scattered light from above the sample. As seen in Fig. 3(j), significant out-of-plane scattering is observed from the random structure. However, Fig. 3(k-l) shows the out-of-plane scattering loss is significantly reduced in the photonic amorphous structure and the golden-angle spiral lattice. This observation confirms that adding structural correlation to the scattering medium can reduce the insertion loss and improve the collection efficiency of the spectrometer. For a quantitative comparison, we estimated the transmission for the three spectrometers, by dividing the sum of the intensities of all detectors by the intensity of the scattered light at the entrance of the spectrometer. The transmission was measured in the wavelength range of 1500 nm – 1525 nm with a 0.25 nm step and then averaged. The amorphous and spiral structures exhibited 2.85 and 2.77 times higher transmission than the random structure, respectively. In addition to reducing the out-of-plane scattering, the amorphous and spiral spectrometers maintained similar spectral resolution and bandwidth to the random spectrometer (see Supplementary Information).

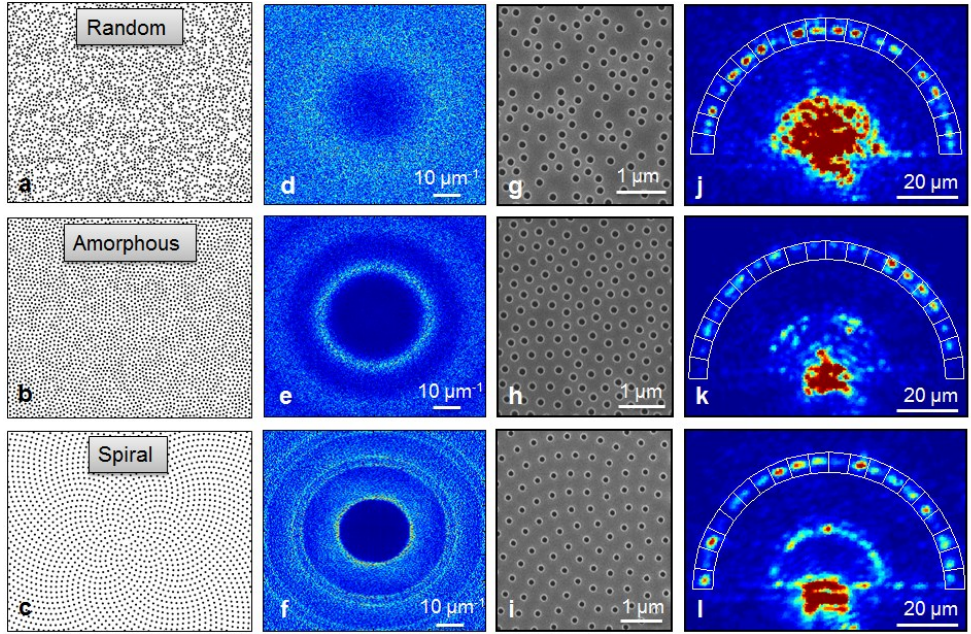

**Figure 3 | Amorphous and spiral spectrometers with reduced out-of-plane leakage**. Real space patterns (a-c), 2D Fourier spectra (d-f) and SEM images (g-i) of the random structure, photonic amorphous structure, and golden-angle spiral lattice used for on-chip spectrometers. While the random structure has all spatial vectors, the amorphous and spiral structures have dominant spatial vectors represented by the bright circles due to structural correlations. (j-l) Optical near-field images of the three spectrometers. The input wavelength is set at $\lambda = 1500$ nm. The intensity of light scattered out-of-plane is dramatically reduced in the amorphous and spiral structures.

In summary, we have utilized multiple scattering in a disordered photonic structure to realize an on-chip spectrometer. The enhanced optical pathlength afforded by multiple scattering enables high resolution with a small footprint. A photonic crystal boundary was used to confine the light in the disordered medium and channel it to the detectors. The input spectra were accurately reconstructed from the spatial intensity distributions of transmitted light. We achieved 0.75 nm resolution with 25 nm bandwidth around the wavelength of 1500 nm with a semicircular random structure of 25 μm radius. Finally, we engineered the disorder to reduce the out-of-plane scattering loss. By replacing the completely random structure with a photonic amorphous structure or a golden-angle spiral lattice, we were able to control the spatial vectors available for out-of-plane scattering. Such a high resolution, compact on-chip spectrometer could enable lab-on-a-chip spectroscopy applications.

**Methods**

The spectrometers were fabricated on silicon-on-insulator (SOI) wafers with a 220 nm silicon layer on top of a 3 μm $SiO_2$ layer. The random structure, photonic crystal boundary, and the coupling waveguides were all defined during a single electron-beam lithography exposure. The pattern was then transferred to the silicon layer via reactive ion etching in a Chlorine environment. The scattering media (random, amorphous, and spiral) consisted of 75 nm radius air cylinders. The photonic crystal boundaries were designed to support a full bandgap for TE polarized light in the wavelength range of 1478 nm -1560 nm. They are formed by triangular arrays of 180 nm radius air holes with a lattice constant of 505 nm. The spectrometer was tested using a tunable, near-IR laser (HP 8168F) which was coupled to a single-mode, polarization maintaining lensed fiber. The lensed fiber delivered the laser beam to the ridge waveguide at the cleaved edge of the chip. The tunable laser was used to calibrate the spectrometer transmission matrix and then to test the ability of the spectrometer to reconstruct various probe spectra. The device was tested under TE polarization (electric field in the plane of the wafer). Scattered light was imaged from above the chip using a 50× objective (NA=0.55) and an InGaAs camera (Xenics Xeva 1.7-320). The random spectrometer also works for TM polarized light provided that the transmission matrix for the TM polarization is calibrated.

**Acknowledgements**

We acknowledge Aristide Dogariu, Mathis Fink, Allard Mosk, Alexey Yamilov, and Sylvain Gigan for useful discussions. This work was supported by National Science Foundation under grants Nos. DMR-1205307 and ECCS-1128542. Computational resources were provided under the Extreme Science and Engineering Discovery Environment (XSEDE) grant No. DMR-100030. Facilities use was supported by YINQE and NSF MRSEC DMR-1119826.

**Author contributions** H.C. and B.R. designed the spectrometers. B.R. fabricated the spectormeters and did all the testing and spectral reconstruction. S.F.L. performed the FDFD simulation of spectrometers, and R.S. helped B.R. characterize the spectral correlation of speckle patterns in random media. B.R. and H.C. prepared the manuscript with the input from S.F.L.

**Additional information** Supplementary information is available in the online version of the paper. Reprints and permissions information is available online at www.nature.com/reprints. Correspondence and requests for materials should be addressed to B.R. and H.C.

**Competing Interests** The authors declare that they have no competing financial interests.